\documentclass[11pt,reqno]{amsart}

\usepackage{amscd,amssymb,amsmath,amsthm}
\usepackage[arrow,matrix]{xy}
\usepackage{graphicx}
\usepackage{epstopdf}
\usepackage{color}
\usepackage{cite}
\newtheorem{thm}{Theorem}
\newtheorem{lemma}{Lemma}
\newtheorem{pro}{Proposition}

\newtheorem{rk}{Remark}
\newtheorem{defn}{Definition}

\newcommand{\bc}{\begin{center}}
\newcommand{\ec}{\end{center}}
\newcommand{\be}{\begin{equation}}
\newcommand{\ee}{\end{equation}}
\newcommand{\bea}{\begin{eqnarray}}
\newcommand{\eea}{\end{eqnarray}}
\newcommand{\ba}{\begin{array}}
\newcommand{\ea}{\end{array}}
\newcommand{\edc}{\end{document}}

\begin{document}
\sloppy

ÓÄÊ 517.98
\begin{center}
\textbf{\Large {Weakly periodic Gibbs measures for two and three state HC models on a Cayley tree}}\\
\end{center}

\begin{center}
R.M.Khakimov\footnote{Namangan State University, 316, Uychi str., 160119, Namangan, Uzbekistan.\\
E-mail: rustam-7102@rambler.ru}, G.T.Madgoziyev\footnote{Academic lyceym N2 under Tashkent University of Information Technologies, 112, Bogishamol str., 100084, Tashkent, Uzbekistan.\\
E-mail: gmadgoziyev@yandex.ru},
\end{center}\

In this paper we study two and three state HC-model on a Cayley tree. Under some conditions on
parameters of the HC-model, in the case of normal divisor
of the index four, we prove the existence of the
weakly periodic (non periodic) Gibbs measures which
are different from the known weakly periodic measures (see \cite{KhR}).
Moreover, we find some regions for the $\lambda$ parameter ensuring that
a extreme measure is not unique.\

\textbf{Keywords}: Cayley tree, configuration, HC-model, Gibbs
measure, periodic measure, weakly periodic measure.\

\section{Introduction}

The description of all limiting Gibbs measures for a given
Hamiltonian is one of main problems in the theory of Gibbs
measures. The Gibbs measure is a fundamental law determining
the probability of a microscopic state of a given physical system.
Each Gibbs measure is associated with one phase of the physical system,
and if a Gibbs measure is non unique, then it is said that there is a phase
transition \cite{6}-\cite{Si}. The study of Gibbs measures therefore plays an important
role in many fields of science, in particular, in statistical mechanics,
physics, biology, and queuing theory \cite{Ba}, \cite{Kel1}.
It is well known that the set of all limit Gibbs measures forms a nonempty convex compact subset of the set of all probability measures \cite{6}-\cite{Si} and each point (i.e., Gibbs measure) of this convex set can be uniquely expanded in its extreme points. In this connection, it is especially interesting to describe all extreme points of this convex set, i.e., the extreme Gibbs measures.
The reader can find the definitions and of other
subjects related to the theory of Gibbs measures, for example,
in \cite{6}-\cite{Si}. In \cite{Rb} the theory of Gibbs measures on
Cayley trees is presented.

In \cite{7} a HC (Hard Core) model with two states on a Cayley
tree was studied and it was proved that the translation-invariant
Gibbs measure is unique for this model. Moreover, under certain
conditions on the HC-model parameters, it was proved the
non uniqueness of periodic Gibbs measures with the period two.

In \cite{8}, \cite{9} the notion of the periodic Gibbs measure
was extended to a more general
notion called weakly periodic Gibbs measure,
where the authors have constructed such measures for the Ising model on the Cayley
tree.

In \cite{RKh}, weakly periodic Gibbs measures were studied for the HC-model with respect to
normal divisors of index two of group representation of the Cayley tree.
Under certain conditions imposed on the parameters, it was shown that the weakly periodic Gibbs measure is
unique (translation invariant). In \cite{XR} the uniqueness of the weakly periodic Gibbs measure for the
HC-model with arbitrary values of the parameters is proved.

The work \cite{KhR} is devoted to the investigation of weakly periodic Gibbs measures for the HC-model with
respect to a normal divisor of index four on a Cayley tree. The existence of weakly periodic (non periodic) Gibbs
measures on some invariants under certain conditions imposed on the parameters is proved.

Gibbs measures for the 3-state HC-models on the Cayley tree of order $k\geq1$ were studied in \cite{BW}-\cite{RKh1}.
The translation-invariant Gibbs measure is shown to be not unique. The fertile HC-models are pointed
out in \cite{BW} corresponding to the hinge, pipe, wand, and wrench graphs in the 3-state case.
The work \cite{RKh1} is devoted to study translation-invariant Gibbs measure for three state HC models on a Cayley tree of order $k$. On a Cayley tree of order two regions in which translation-invariant Gibbs measures is extreme or non-extreme in the set of all Gibbs measures was found.

In this paper, under some conditions on parameters of the HC-model, we prove existence of the
weakly periodic (non periodic) Gibbs measures for a normal divisor
of the index four, which are different from the weakly periodic measures of the work \cite{KhR}.
Moreover, we find some regions for the $\lambda$ parameter ensuring that
a extreme measure is not unique.

\section{Preliminary Information.}

A Cayley tree $\tau^k$ of order $ k\geq 1 $ is an infinite tree, i.e., a graph without cycles such that each
vertex has precisely $k+1$ edges.

Let $ \tau^k=(V,L,i)$, where $V$ is the set of vertices of the graph $ \tau^k$, $L$
is the set of its edges, and $i$ is the incidence function associating each edge $l\in L$ to its endpoints $x, y \in
V$. If $i (l) = \{ x, y \} $, then $x$ and $y$ are
called the nearest neighbors of a vertex, and we write this as $l = \langle
x,y\rangle $. The distance $d(x,y), x, y \in V$ on
the Cayley tree is defined as
$$
d (x, y) = \min \{d | \exists x=x_0,x_1, \dots, x_{d-1},
x_d=y\in V \ \ \mbox {such, that} \ \ \langle x_0,x_1\rangle,\dots, \langle x_
{d-1}, x_d\rangle\}.$$

For a fixed $x^0\in V$ we set
$$ W_n = \ \{x\in V\ \ | \
\ d (x, x^0) =n \}, \  V_n = \ \{x\in V\ \ | \ \ d (x, x^0) \leq n
\}$$

For $x\in W_{n}$ we denote
$ S(x)=\{y\in{W_{n+1}}:d(x,y)=1\}.$

Consider the $m$-state nearest-neighbor HC-model on the Cayley tree. In this model to each vertex
$x$ there corresponds one of the values $\sigma (x)\in \Phi=\{0,1,..,m\}$. The values $\sigma (x)=1,..,m$
indicate that the vertex $x$ is occupied, while $\sigma (x)=0$ indicates that $x$ is vacant.

A configuration $\sigma=\{\sigma(x),\ x\in V\}$ on the Cayley tree is defined as a function from $V$ to $\Phi$. Denote by $\Omega$ the set of all configurations on $V$. Similarly, we can define configurations in $V_n$ and $W_n$, and denote by $\Omega_{V_n}$ and $\Omega_{W_n}$ the sets of all configurations in $V_n$ and $W_n$.

Regard $\Phi$ as the vertex set of some graph $G$. Using $G$, define some $G$-admissible configuration as
follows: A configuration $\sigma$ is called a $G$-admissible configuration on the Cayley tree (in $V_n$ or $W_n$) whenever $\{\sigma (x),\sigma (y)\}$ is the edge of $G$ for every pair $x,y$ of nearest neighbors in $V$ ($V_n$). Denote by $\Omega^G$ ($\Omega_{V_n}^G$) the sets of $G$-admissible configurations.

An activity set \cite{BW} for a graph $G$ is a function $\lambda:G
\to R_+$ from the vertex set of $G$ into the set of positive real numbers. The value $\lambda_i$ of $\lambda$ at $i\in\{0,1,..,m\}$ is called an \emph{activity} of $G$.

Given $G$ and $\lambda$, define the Hamiltonian of the $G$-HC-model as

$$H^{\lambda}_{G}(\sigma)=\left\{%
\begin{array}{ll}
     \sum\limits_{x\in{V}}{\log \lambda_{\sigma(x)},} \ \ \ $ if $ \sigma \in\Omega^{G} $,$ \\
   +\infty ,\ \ \ \ \ \ \ \ \ \  \ \ \ $  \ if $ \sigma \ \notin \Omega^{G} $.$ \\
\end{array}%
\right. $$

\begin{defn}\label{d0} \cite{BW}.
A graph is called \emph{fertile} if there exists an activity tuple $\lambda$ such that the corresponding Hamiltonian has at least two translation-invariant Gibbs measures.
\end{defn}

This article considers the cases $m =1$ and $m =2$. For $m =2$ we consider the case $\lambda_0=1$,
$\lambda_1=\lambda_2=\lambda$, and study the corresponding translation-invariant Gibbs measures, while for $m =1$,
the weakly periodic Gibbs measures.

For the 3-state HC-model we consider two types of fertile graphs with three vertices $0, 1, 2$ (on the
set of values of $\sigma(x)$), which are of the following form: the hinge $\{0,0\}, \{0,1\}, \{0,2\}, \{1,1\}, \{2,2\}$; the wand $\{0,1\}, \{0,2\}, \{1,1\}, \{2,2\}$ (see Fig.1). In the case of the 2-state HC-model we take as $G$, the pipe $\{0,0\}, \{0,1\}$ (see Fig.2).
\begin{center}
\includegraphics[width=7cm]{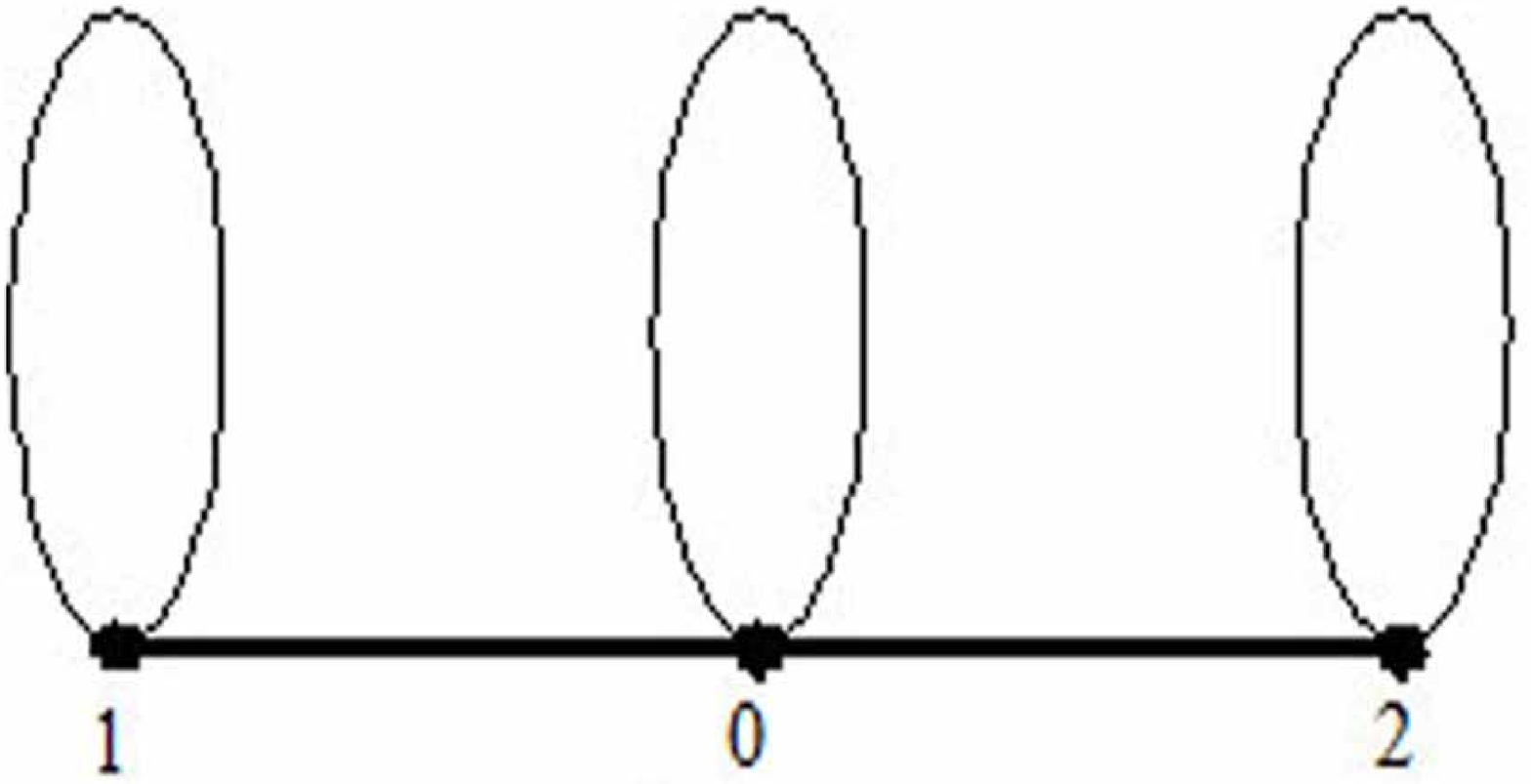} \ \ \ \  \includegraphics[width=7cm]{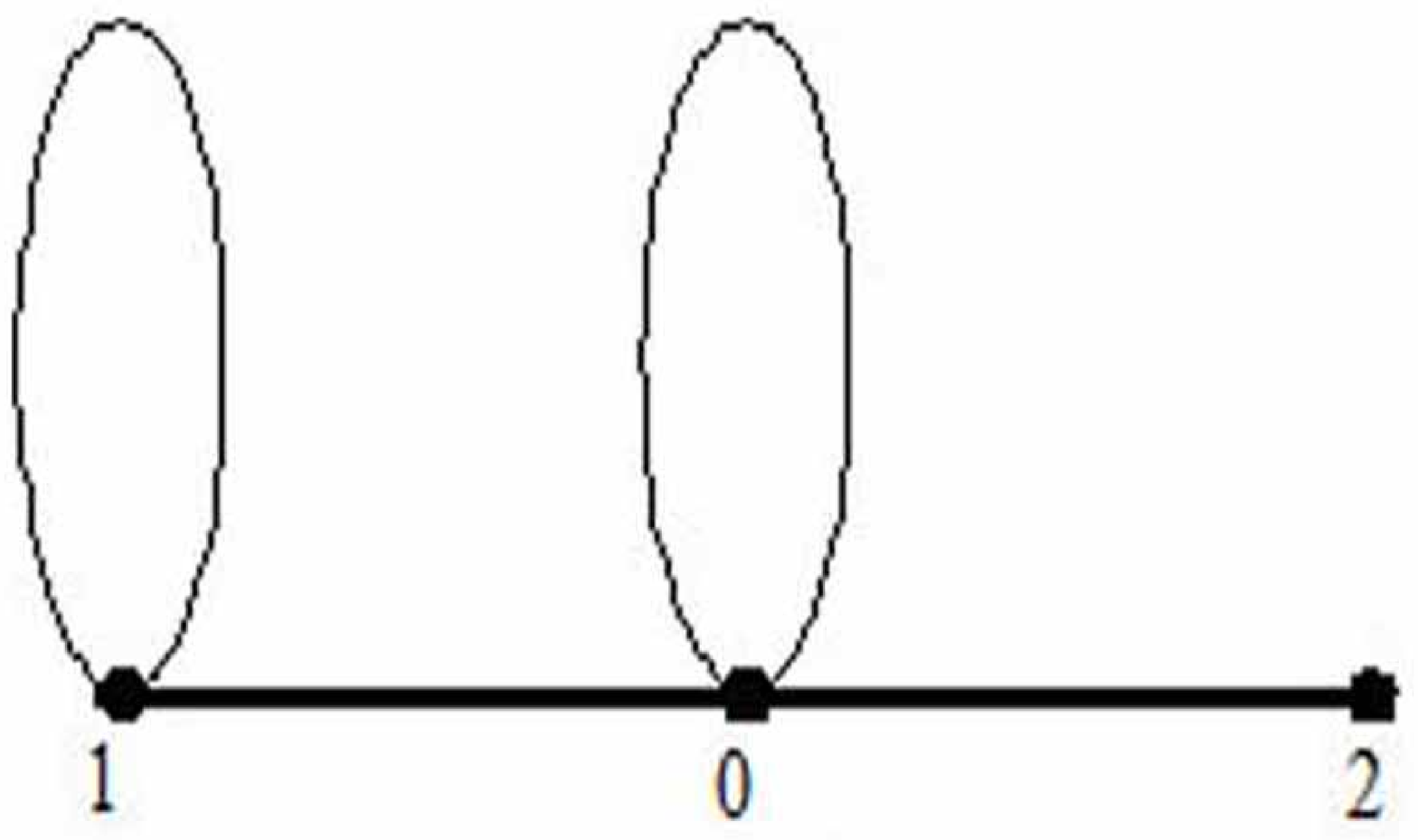} 
\end{center}
\begin{center}{\footnotesize \noindent
 Fig.~1.
  Graphs: the hinge (on the left) and the wand (on the right)} 
\end{center}

\begin{center}
\includegraphics[width=7cm]{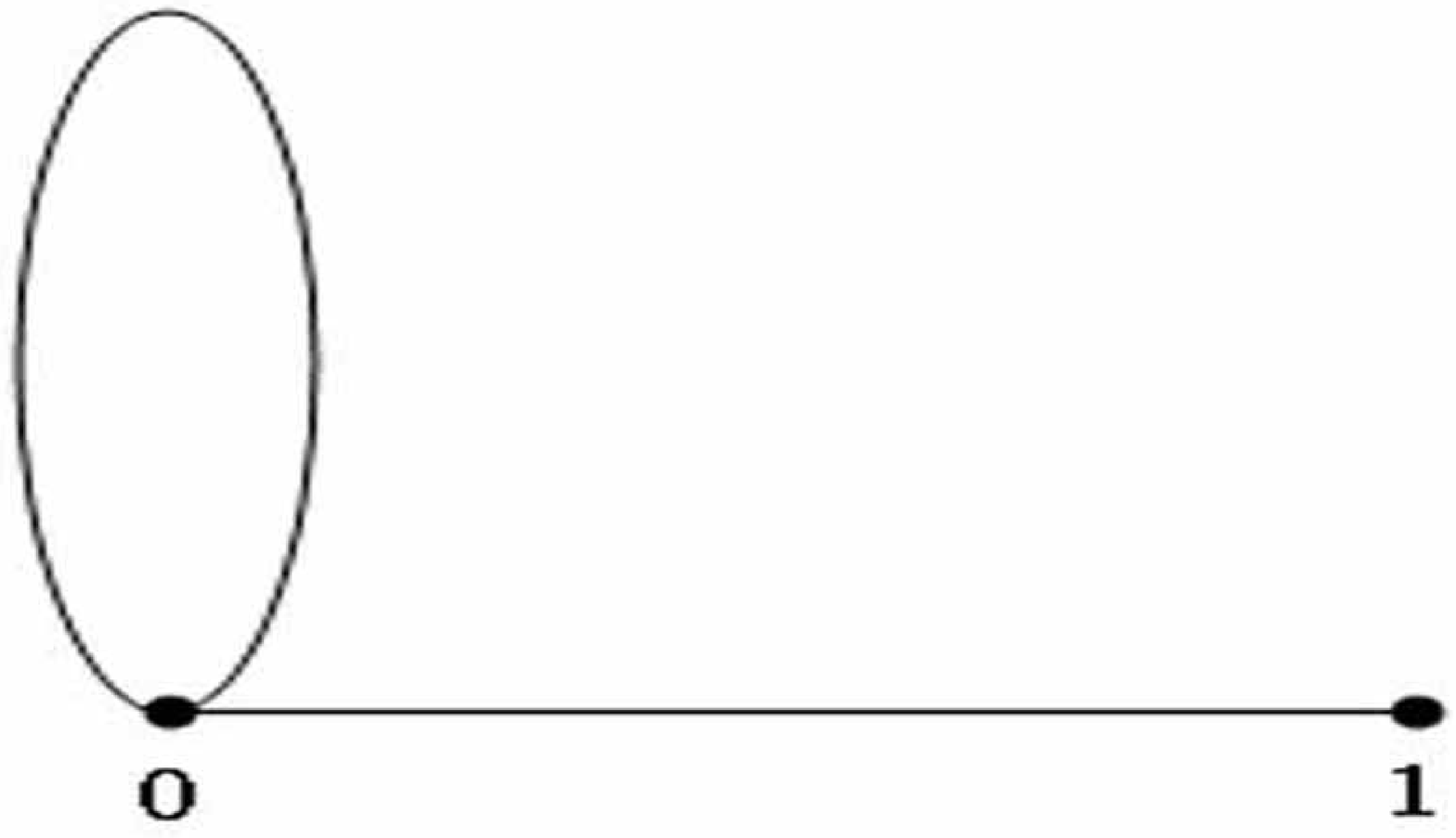}
\end{center}
\begin{center}{\footnotesize \noindent
 Fig.~2.
 The graph: the pipe}
\end{center}
For $\sigma_n\in\Omega_{V_n}$ we let
$$\#\sigma_n=\sum\limits_{x\in V_n}{\mathbf 1}(\sigma_n(x)\geq 1)$$
denote the number of occupied vertices in $\sigma_n$.

Let $z:x\mapsto z_x=(z_{0,x}, z_{1,x}) \in R^2_+$ be a vector-valued function on $V$. For $n=1,2,\ldots$ and $\lambda>0$ we consider the probability measure $\mu^{(n)}$ on $\Omega_{V_n}$
defined as
\begin{equation}\label{e0}
\mu^{(n)}(\sigma_n)=\frac{1}{Z_n}\lambda^{\#\sigma_n} \prod_{x\in
W_n}z_{\sigma(x),x}.
\end{equation}
Here $Z_n$ is a normalization factor,
$$
Z_n=\sum_{{\widetilde\sigma}_n\in\Omega^G_{V_n}}
\lambda^{\#{\widetilde\sigma}_n}\prod_{x\in W_n}
z_{{\widetilde\sigma}(x),x}.
$$
The probability measure $\mu^{(n)}$ is said to be consistent if for all $n\geq 1$ and any
$\sigma_{n-1}\in\Omega_{V_{n-1}}$:
\begin{equation}\label{e00}
\sum_{\omega_n\in\Omega_{W_n}}
\mu^{(n)}(\sigma_{n-1}\vee\omega_n){\mathbf 1}(
\sigma_{n-1}\vee\omega_n\in\Omega^G_{V_n})=
\mu^{(n-1)}(\sigma_{n-1}).
\end{equation}
In this case, there is a unique measure $\mu$ on $(\Omega,
\textbf{B})$ such that
$$\mu(\{\sigma|_{V_n}=\sigma_n\})=\mu^{(n)}(\sigma_n),$$
for all $n$ and any $\sigma_n\in
\Omega_{V_n}$, where $\textbf{B}$ is the $\sigma$-algebra generated by cylindrical subsets of the set $\Omega$.

\begin{defn}\label{d00} The measure $\mu$ defined in (\ref{e0}) under condition (\ref{e00}) is called a ($G$)-HC-Gibbs measure with $\lambda>0$ corresponding to the function $z:\,x\in V
\setminus\{x^0\}\mapsto z_x$. In this case, a HC Gibbs measure corresponding to a constant
function $z_x\equiv z$ is said to be translation-invariant.
\end{defn}

\section{Weakly periodic Gibbs measures}

It is known that there is a one-to-one correspondence between the set $V$ of vertices of the Cayley tree
of order $ k\geq 1$ and the group $G_k$ which is the free product of $k+1$  cyclic groups of order two with the
generators $a_1,...,a_{k+1}$ \cite{1}, \cite{2}.

It is known \cite{7} that each Gibbs measure for the HC-model on a Cayley tree can be associated with a collection of quantities $z=\{z_x, x\in G_k \}$ satisfying the equality
\begin{equation}\label{e1}
z_x=\prod_{y \in S(x)}(1+\lambda z_y)^{-1},
\end{equation}
where $\lambda=e^{J\beta}>0$ is a parameter, $\beta={1\over T}$ and $T>0$ is temperature.

Let $\widehat{G}_k$ be a subgroup of the group $G_k$.

\begin{defn}\label{d1} A collection of quantities $z=\{z_x,x\in G_k\}$
is called $ \widehat{G}_k$-periodic if  $z_{yx}=z_x$ for any
$x\in G_k$ and $y\in\widehat{G}_k.$\
\end{defn}
The $G_k$-periodic collections are called translation invariant.

For any $x\in G_k $ the set $\{y\in G_k: \langle
x,y\rangle\}\setminus S(x)$ contains a unique element denoted by $x_{\downarrow}$ (see \cite{8}).

Let $G_k/\widehat{G}_k=\{H_1,...,H_r\}$ be a quotient group, where
$\widehat{G}_k$ is a normal divisor of index $r\geq 1.$\

\begin{defn}\label{d2} A collection of quantities $z=\{z_x,x\in G_k\}$
is called $\widehat{G}_k$-weakly periodic if
$z_x=z_{ij}$ for $x\in H_i, x_{\downarrow}\in H_j$ for any $x\in G_k$.\
\end{defn}
Note that a weakly periodic collection $z$ coincides with the ordinary periodic collection (see Definition \ref{d1}) if the value of $z_x$ is independent of $x_{\downarrow}$.\

\begin{defn}\label{d3} A measure $\mu$ is called
$\widehat{G}_k$-(weakly) periodic if it corresponds to a
$\widehat{G}_k$-(weakly) periodic collection of quantities $z$.\
\end{defn}

Let $A\subset\{1,2,...,k+1\}$ and let
$H_A=\{x\in G_k:\sum\limits_{i\in A}w_x(a_i) \ \mbox{is an even number} \},$
where $w_x(a_i)$ is the number of letter $a_i$ in the word $x\in G_k$,
$$G_k^{(2)}=\{x\in G_k: \ \ \mid x\mid \mbox{is an even number}\},$$
where $\mid x\mid$ is the length of the word $x\in
G_k$, and $G_k^{(4)}=H_A\cap G_k^{(2)}$ is the normal divisor of index four.\

Consider the quotient group $G_k/G_k^{(4)}=\{H_0, H_1, H_2, H_3\},$ where
$$H_0=\{x\in G_k: \sum\limits_{i\in
A}w_x(a_i) \ \mbox{is even}, |x| \ \mbox{is even}\},$$
$$H_1=\{x\in G_k: \sum\limits_{i\in
A}w_x(a_i) \ \mbox{is odd}, |x| \ \mbox{is even}\},$$
$$H_2=\{x\in G_k: \sum\limits_{i\in
A}w_x(a_i) \ \mbox{is even}, |x| \ \mbox{is odd}\},$$
$$H_3=\{x\in G_k: \sum\limits_{i\in
A}w_x(a_i) \ \mbox{is odd}, |x| \ \mbox{is odd}\}.$$
Thus by virtue of (\ref{e1}), the $G_k^{(4)}-$weakly periodic collection of quantities $z_x$ has the form
$$
z_x=\left\{%
\begin{array}{ll}
    z_1, & {x \in H_3, \ x_{\downarrow} \in H_1} \\
    z_2, & {x \in H_1, \ x_{\downarrow} \in H_3} \\
    z_3, & {x \in H_3, \ x_{\downarrow} \in H_0} \\
    z_4, & {x \in H_0, \ x_{\downarrow} \in H_3} \\
    z_5, & {x \in H_1, \ x_{\downarrow} \in H_2} \\
    z_6, & {x \in H_2, \ x_{\downarrow} \in H_1} \\
    z_7, & {x \in H_2, \ x_{\downarrow} \in H_0} \\
    z_8, & {x \in H_0, \ x_{\downarrow} \in H_2}, \\
    \end{array}%
\right.$$
where $z_i$, $i=1,\dots,8$ satisfy system of equations
\begin{equation}\label{e2}
\left\{%
\begin{array}{ll}
    z_{1}=\frac{1}{(1+\lambda z_4)^i}\cdot\frac{1}{(1+\lambda z_2)^{k-i}} \\
    z_{2}=\frac{1}{(1+\lambda z_6)^i}\cdot\frac{1}{(1+\lambda z_1)^{k-i}} \\
    z_{3}=\frac{1}{(1+\lambda z_4)^{i-1}}\cdot\frac{1}{(1+\lambda z_2)^{k-i+1}}\\
    z_{4}=\frac{1}{(1+\lambda z_3)^{i-1}}\cdot\frac{1}{(1+\lambda z_7)^{k-i+1}}\\
    z_{5}=\frac{1}{(1+\lambda z_6)^{i-1}}\cdot\frac{1}{(1+\lambda z_1)^{k-i+1}}\\
    z_{6}=\frac{1}{(1+\lambda z_5)^{i-1}}\cdot\frac{1}{(1+\lambda z_8)^{k-i+1}}\\
    z_{7}=\frac{1}{(1+\lambda z_5)^i}\cdot\frac{1}{(1+\lambda z_8)^{k-i}} \\
    z_{8}=\frac{1}{(1+\lambda z_3)^i}\cdot\frac{1}{(1+\lambda z_7)^{k-i}}. \\
\end{array}%
\right.
\end{equation}
Here $i=|A|$ is the cardinality of the set $A$.

By (\ref{e2}), after some transformations, we obtain the following system of equations (see \cite{KhR}):
\begin{equation}\label{e3}
\left\{%
\begin{array}{ll}
    z_{1}=\frac{(1+\lambda z_{7})^k}{((1+\lambda z_7)^{k/i}+\lambda z_8^{1-1/i})^i}\cdot\frac{1}{(1+\lambda z_2)^{k-i}} \\
    \\
    z_{2}=\frac{(1+\lambda z_8)^k}{((1+\lambda z_8)^{k/i}+\lambda z_7^{1-1/i})^i}\cdot\frac{1}{(1+\lambda z_1)^{k-i}} \\
    \\
    z_{7}=\frac{(1+\lambda z_1)^k}{((1+\lambda z_1)^{k/i}+\lambda z_2^{1-1/i})^i}\cdot\frac{1}{(1+\lambda z_8)^{k-i}} \\
    \\
    z_{8}=\frac{(1+\lambda z_2)^k}{((1+\lambda z_2)^{k/i}+\lambda z_1^{1-1/i})^i}\cdot\frac{1}{(1+\lambda z_7)^{k-i}} \\
\end{array}%
\right.
\end{equation}
We consider next invariant sets (see \cite{KhR})
$$I_1=\{(z_1, z_2, z_7, z_8) \in R^4: z_1=z_2=z_7=z_8\}, \ \ I_2=\{(z_1, z_2, z_7, z_8)\in R^4: z_1=z_7, \ z_2=z_8\},$$
$$I_3=\{(z_1, z_2, z_7, z_8) \in R^4: z_1=z_2, z_7=z_8\}, \ \ I_4=\{(z_1, z_2, z_7, z_8)\in R^4: z_1=z_8, \ z_2=z_7\}.$$
The following assertions are known from \cite{KhR}

\begin{lemma}\label{l2} \cite{KhR} \textit{If weakly periodic Gibbs measures exist on the invariant sets $I_2,
I_3, I_4$, then they are either translation-invariant or weakly periodic (nonperiodic).}\
\end{lemma}
\begin{thm} \cite{KhR} \textit{For the HC-model, in the case of normal divisor of index four, the following assertions are true:}

\textit{1. For $k\geq1$ and $i\leq k$, a weakly periodic Gibbs measure is unique on $I_1$. Moreover, this measure
coincides with the unique translation-invariant Gibbs measure.}

\textit{2. Let $k=2$, $i=1$, and $\lambda_{cr}=4$. Then, on $I_2$ there exists one weakly periodic Gibbs measure, which is translation-invariant, for $\lambda<\lambda_{cr}$, two weakly periodic Gibbs measures one of which is translation-invariant and the other is weakly periodic (nonperiodic) for
$\lambda=\lambda_{cr}$, and at most two weakly periodic
(nonperiodic) Gibbs measures for $\lambda>\lambda_{cr}$.}

\textit{3. Let $k=3$ and $i=1$. Then one can find $\lambda_0$ such that, on $I_2$ for $\lambda>\lambda_{0}$, there exist at least four Gibbs measures one of which is translation-invariant and the others are weakly periodic (nonperiodic) Gibbs measures.}

\textit{4. For $k\geq1$ and $i=1$, a weakly periodic Gibbs measure is unique on $I_3$.}

\textit{5. For $k=2,3$ and $i=1$, a weakly periodic Gibbs measure is unique on $I_4$.}
\end{thm}

\textbf{The case $I_2$.} We rewrite system of equations (\ref{e3}) on $I_2$ for $k=2,
i=2$
\begin{equation}\label{e4}
\left\{%
\begin{array}{ll}
    z_{1}=\frac{(1+\lambda z_{1})^2}{(1+\lambda z_1+\lambda\sqrt{z_2})^2}
    \\[2mm]
    z_{2}=\frac{(1+\lambda z_2)^2}{(1+\lambda z_2+\lambda\sqrt{z_1})^2 }. \\
    \end{array}%
\right.
\end{equation}

We denote $s=\sqrt{z_1}$ and $t=\sqrt{z_2}$. Then the system of equations (\ref{e4}) has the form

\begin{equation}\label{e41}
\left\{%
\begin{array}{ll}
    s=\frac{1+\lambda s^2}{1+\lambda s^2+\lambda t}
    \\[2mm]
    t=\frac{1+\lambda t^2}{1+\lambda t^2+\lambda s }. \\
    \end{array}%
\right.
\end{equation}
It is easy to see that $0<s<1, \ 0<t<1$ in (\ref{e41}), i.e. $0<z_1<1, \ 0<z_2<1$ in (\ref{e4}). Consider two cases: $s=t$ and $s\neq t$.

If $s=t$, i.e. $z_1=z_2$ in (\ref{e4}) (this means $z_1=z_2=z_7=z_8$ on $I_2$) then it is easy to show that $z_1=z_2=z_3=z_4=z_5=z_6=z_7=z_8$ in (\ref{e3}).

Consider the case $s\neq t$. In the case from the second equation of system (\ref{e41}) we find $s$ and substitute in the first equation. As a result, we get the following equation
$$\lambda^3t^9-3\lambda^3t^8+(4\lambda^3+3\lambda^2)t^7-(9\lambda^2+2\lambda^3)t^6+
(12\lambda^2+3\lambda)t^5-(7\lambda^2+9\lambda)t^4+$$
$$+(2\lambda^2+11\lambda+1)t^3-(6\lambda+3)t^2+(\lambda+3)t-1=0,$$
which has a solution $t=t(\lambda)$. But we regard this as an equation for $\lambda$ and by using the Cardano formula, we find next solutions
$$\lambda_1=-\frac{(t-1)^2}{(t^2-2t-2)^2}=\varphi_1(t), \ \lambda_2=\frac{1-t}{t^3}=\varphi_2(t), \ \lambda_3=\frac{1}{t(1-t)}=\varphi_3(t).$$
Analysis of functions $\varphi_1(t), \ \varphi_2(t), \ \varphi_3(t)$ shows that $\varphi_1(t)<0$ for any values $t>0$, and $\varphi_2(t)>0, \ \varphi_3(t)>0$ for $0<t<1$.
Moreover $\varphi_2(t)\rightarrow \infty$ for $t\rightarrow0$ and $\varphi_2(t)\rightarrow0$ for $t\rightarrow 1$. Analogically $\varphi_3(t)\rightarrow \infty$ for $t\rightarrow0$ and $t\rightarrow 1$.\

The derivative of the function $\varphi_2(t)$ is negative,
i.e. the function $\varphi_2(t)$ decreases for $0<t<1$. Hence,
each value of $\lambda_2$ corresponds to only one
value of $t$ for $0<t<1$ (see Fig.3).

The derivative $\varphi_3^{'}(t)$ is negative for $0<t<t_{cr}$ and it is positive for $t_{cr}<t<1$, where $t_{cr}=\frac{1}{2}$ is a solution of the equation $\varphi_3^{'}(t)=0$. Hence, the function $\varphi_3(t)$ decreases for $0<t<t_{cr}$ and it increases for $t_{cr}<t<1$.
Moreover, we have $\varphi_3^{''}(t)>0$ for $0<t<1$, i.e. each value of $\lambda_3$ corresponds to exactly two
values of $t$ for $\lambda_3>\lambda_{cr}=\varphi_3(t_{cr})=4$,  one value of $t$ at $\lambda_3=\lambda_{cr}$ and the equation $\lambda_3=\varphi_3(t)$ has no solutions for $\lambda_3<\lambda_{cr}$ (see Fig.3).\

\begin{center}
\includegraphics[width=6cm]{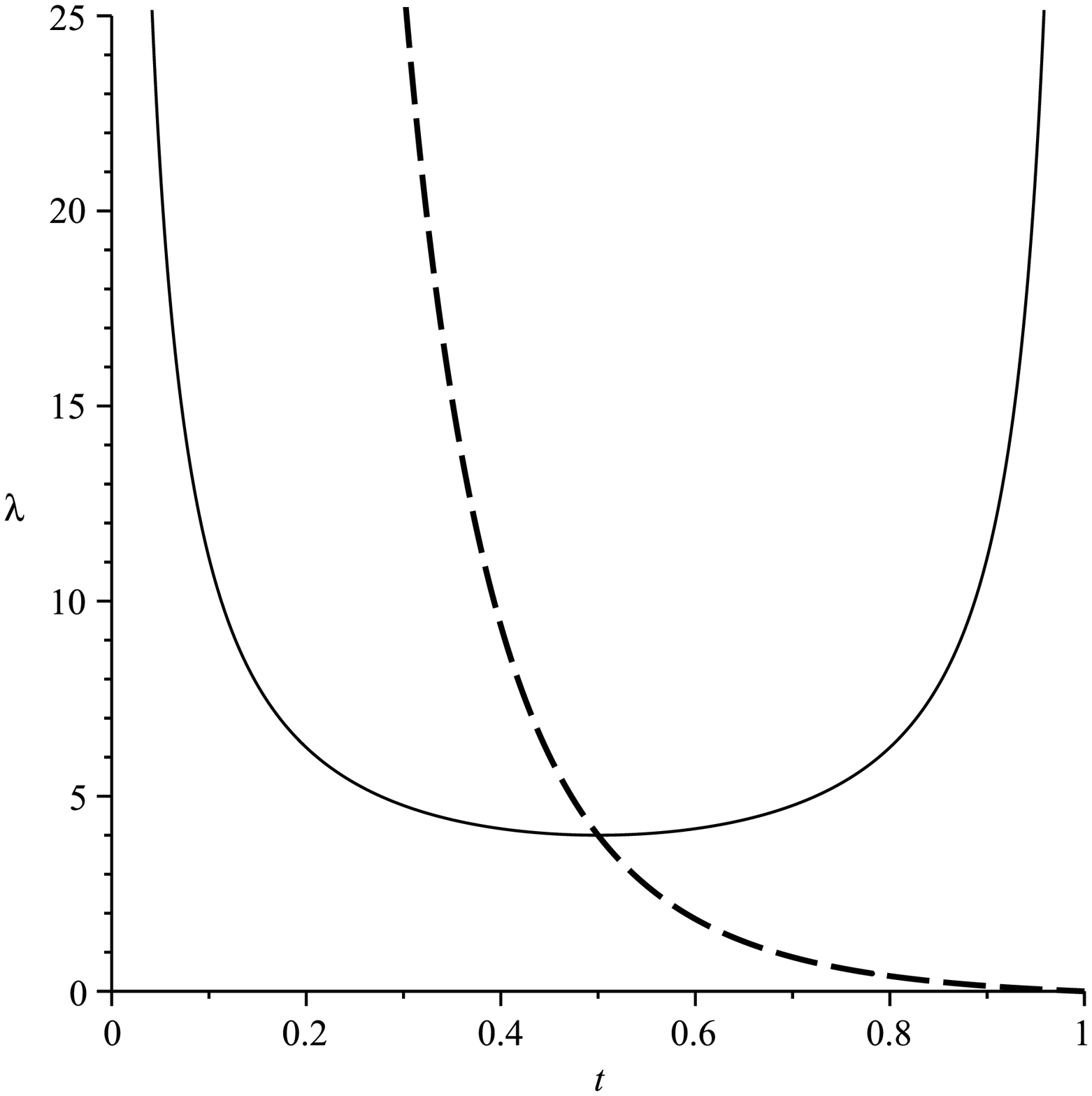}
\end{center}
\begin{center}{\footnotesize \noindent
 Fig.~3.
  Graph of the function $\varphi_2(t)$ (shaded curve) and $\varphi_3(t)$ (continuous curve) for $0<t<1$}.
\end{center}

Thus we proved the following:

\begin{pro}\label{p1} \textit{The system of equations (\ref{e4}) has only one solution for $\lambda<4$, two solutions for  $\lambda=4$ and three solutions for $\lambda>4$.}
\end{pro}

\begin{rk} The solution $\varphi_2(t)$ corresponds to solution (\ref{e41}) for $s=t$.
\end{rk}

 \textbf{The case $I_3$.} The system of equations (\ref{e3}) on $I_3$ for $k=2,
i=2$ has the form
\begin{equation}\label{e7}
\left\{%
\begin{array}{ll}
    z_{1}=\frac{(1+\lambda z_{7})^2}{(1+\lambda z_7+\lambda\sqrt{z_7})^2}
    \\[3mm]
    z_{7}=\frac{(1+\lambda z_1)^2}{(1+\lambda z_1+\lambda\sqrt{z_1})^2 }. \\
    \end{array}%
\right.
\end{equation}
We denote $s=\sqrt{z_1}$ and $t=\sqrt{z_7}$. Then the system of equations (\ref{e7}) has the form

\begin{equation}\label{e71}
\left\{%
\begin{array}{ll}
    s=\frac{1+\lambda t^2}{1+\lambda t^2+\lambda t}
    \\[3mm]
    t=\frac{1+\lambda s^2}{1+\lambda s^2+\lambda s }. \\
    \end{array}%
\right.
\end{equation}
It's clear that $0<s<1, \ 0<t<1$ in (\ref{e71}), i.e. $0<z_1<1, \ 0<z_7<1$ in (\ref{e7}). By analogy with the case $I_2$ we consider two cases: $s=t$ and $s\neq t$.

In the case $s=t$, we obtain
$z_1=z_2=z_3=z_4=z_5=z_6=z_7=z_8.$

In the case $s\neq t$ the expression for s from the first equation in (\ref{e71}) is substituted into the second equation and we obtain an equation with respect to the variable $\lambda$
$$2t^5\lambda^3+(t^5+t^4+3t^3-2t^2)\lambda^2+(2t^3-1)\lambda+t-1=0,$$
which by the Cardano formula has solutions of the form
$$\tilde{\lambda}_1(t)=\frac{1-t}{t^3}, \ \tilde{\lambda}_2(t)=-\frac{t^2+t+1-\sqrt{t^4+2t^3-5t^2+2t+1}}{4t^2},$$
$$\tilde{\lambda}_3(t)=-\frac{t^2+t+1+\sqrt{t^4+2t^3-5t^2+2t+1}}{4t^2}.$$

We note that $\tilde{\lambda}_1(t)>0$ for $0<t<1$ and
$\tilde{\lambda}_2(t)<0, \ \tilde{\lambda}_3(t)<0$ for $t>0$.
Therefore, from the above analysis for the function $\tilde{\lambda}_1(t)$ we have the following

\begin{pro}\label{p2} \textit{The system of equations
(\ref{e7}) has a unique solution for $\lambda>0$.}\
\end{pro}
\begin{rk} The solution $\tilde{\lambda}_1(t)$ corresponds to solution (\ref{e71}) for $s=t$ (see Fig. 4).
\end{rk}

\begin{center}
\includegraphics[width=6cm]{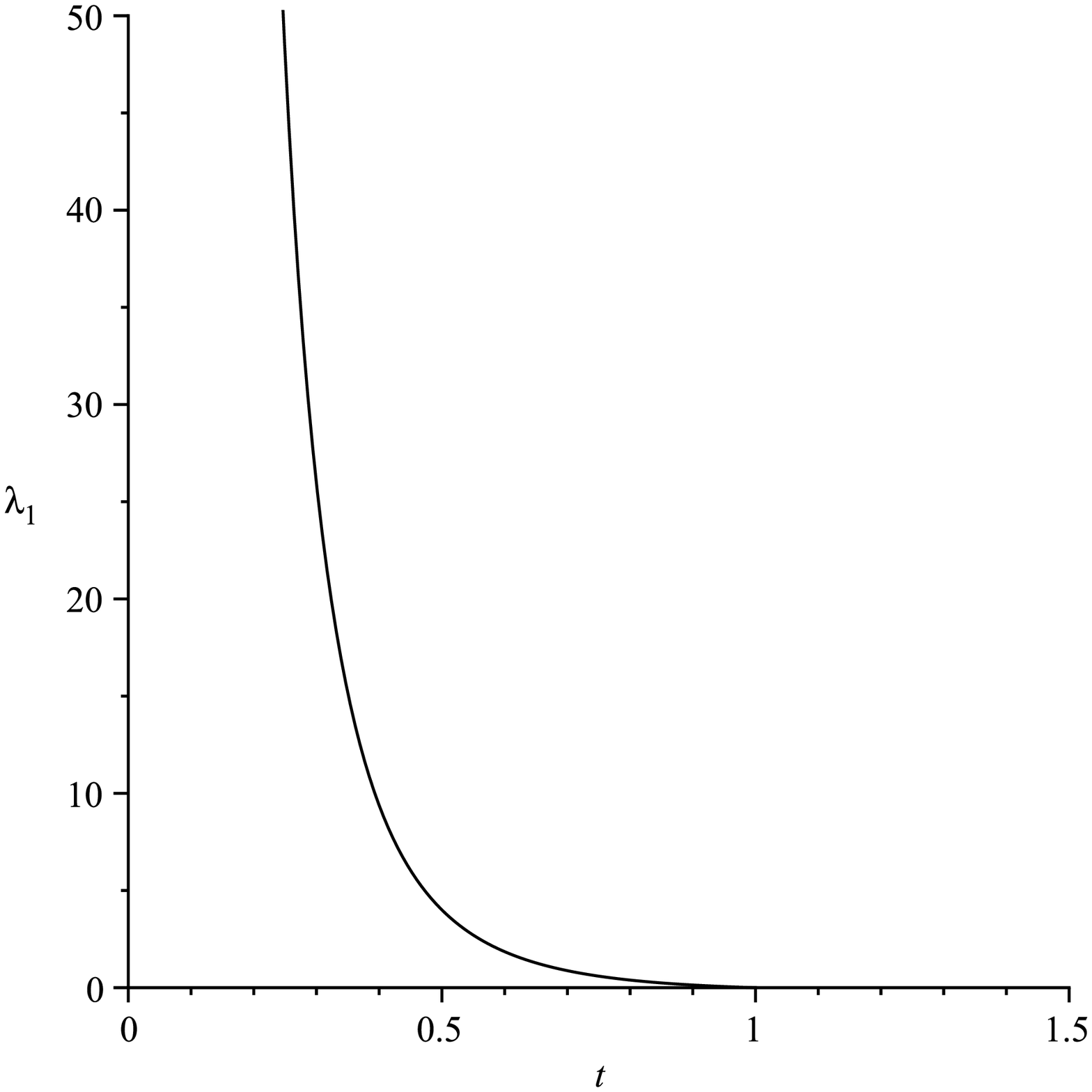}
\end{center}
\begin{center}{\footnotesize \noindent
 Fig.~4.
  Graph of the function $\tilde{\lambda}_1(t)$ for $0<t<1$}.
\end{center}

\textbf{The case $I_4$.} We consider the system of equations (\ref{e3}) on $I_4$:
\begin{equation}\label{e80}
\left\{%
\begin{array}{ll}
    z_{1}=\left(\frac{1+\lambda z_2}{(1+\lambda z_2)^{{k\over i}}+\lambda z_1^{1-{1\over i}}}\right)^i \\[4mm]
    z_{2}=\left(\frac{1+\lambda z_1}{(1+\lambda z_1)^{{k\over i}}+\lambda z_2^{1-{1\over i}}}\right)^i. \\
    \end{array}%
\right.
\end{equation}

Let $k=i$. We denote $\sqrt[k]{z_{1}}=x, \ \sqrt[k]{z_{2}}=y$ and note that $0<z_1<1, \ 0<z_2<1$. Then after some transformations, the last system of equations has the following form

 \begin{equation}\label{e8}
\left\{%
\begin{array}{ll}
    \lambda x^k+\lambda xy^k+x-\lambda y^k=1
    \\[3mm]
     \lambda y^k+\lambda yx^k+y-\lambda x^k=1. \\
    \end{array}%
\right.
\end{equation}

Subtracting from the first equation of system (\ref{e8}) the second we obtain the equation
 \begin{equation}\label{e81}
 (x-y)\cdot[1+\lambda(x^{k-1}+x^{k-2}y+\ldots+xy^{k-2}+y^{k-1})(2-xy)]=0.
 \end{equation}
If $x=y$ then we have $z_1=z_2=\ldots =z_7=z_8$ and a solution of this kind exists and is unique for any values $\lambda>0$. In this case from (\ref{e8}) we have
$$\lambda={1-x\over x^{k+1}}=\varphi(x)$$
and it is easy to show that each value $\lambda$ corresponds only one value $x$. Indeed, the derivative
$\varphi'(x)<0$ for $k\geq2$, $0<x<1$, i.e., in this case the function $\varphi (x)$ decreases and $\varphi''(x)>0$. Hence, each value of $\lambda$ corresponds to only one
value of $x$ for $0<x<1$.

We consider the case $x\neq y$. In this case from (\ref{e81}) we have the equation
$$1+\lambda(x^{k-1}+x^{k-2}y+\ldots+xy^{k-2}+y^{k-1})(2-xy)=0,$$
which has no solutions $(x,y), \ 0<x<1, \ 0<y<1$ for any values $\lambda>0$.

So the following is true

\begin{pro}\label{p3} \textit{The system of equations
(\ref{e80}) has a unique solution for $k=i$ and $\lambda>0$.}\
\end{pro}

Propositions \ref{p1}, \ref{p2} and \ref{p3} imply the following

\begin{thm} \textit{For the HC-model, in the case of normal divisor of index four, the following assertions are true:}

\textit{1. Let $k=2$, $i=2$ and $\lambda_{cr}=4$. Then, on $I_2$ there exists one weakly periodic Gibbs measure, which is translation-invariant, for $\lambda<\lambda_{cr}$, two weakly periodic Gibbs measures one of which is translation-invariant and the other is weakly periodic (nonperiodic) for
$\lambda=\lambda_{cr}$, and three weakly periodic
(nonperiodic) Gibbs measures one of which is translation-invariant and others are weakly periodic (nonperiodic) for
$\lambda>\lambda_{cr}$.}

\textit{2. For $k=2$ and $i=2$, a weakly periodic Gibbs measure is unique on $I_3$. Moreover, this measure
coincides with the unique translation- invariant Gibbs measure.}

\textit{3. For $k=i$, a weakly periodic Gibbs measure is unique on $I_4$. Moreover, this measure
coincides with the unique translation- invariant Gibbs measure.}
\end{thm}
\begin{rk} In each subsection of this theorem, the translation-invariant Gibbs measure corresponds to a solution of the form $z_1=z_2=z_3=z_4=z_5=z_6=z_7=z_8$  of the system of equations (\ref{e3}).
\end{rk}

\section{Translation-invariant Gibbs measures}

We consider three state HC models.

Let $L(G)$ be the set of edges of $G$. We let $A\equiv
A^G=\big(a_{ij}\big)_{i,j=0,1,2}$ denote the incidence matrix of
$G$, i.e.
$$ a_{ij}\equiv a^G_{ij}=\left\{\begin{array}{ll}
1,\ \ \mbox{if}\ \ \{i,j\}\in L(G),\\
0, \ \ \mbox{if} \ \  \{i,j\}\notin L(G).
\end{array}\right.$$

The next theorem states a condition on $z_x$ that guarantees that
the measure $\mu^{(n)}$ is consistent.

\begin{thm}\cite{RSh} Probability measures
$\mu^{(n)}$, $n=1,2,\ldots$, defined by (\ref{e0}), are
consistent if and only if the equalities
$$
z'_{1,x}=\lambda \prod_{y\in S(x)}{a_{10}+
a_{11}z'_{1,y}+a_{12}z'_{2,y}\over
a_{00}+a_{01}z'_{1,y}+a_{02}z'_{2,y}},\\[4mm]
z'_{2,x}=\lambda \prod_{y\in S(x)}{a_{20}+
a_{21}z'_{1,y}+a_{22}z'_{2,y}\over
a_{00}+a_{01}z'_{1,y}+a_{02}z'_{2,y}},$$ (where
$z'_{i,x}=\lambda z_{i,x}/z_{0,x}, \ \ i=1,2$) hold for any
$x\in V$.
\end{thm}
We assume that $z_{0,x}\equiv 1$ and $z_{i,x}=z'_{i,x}>0,\ \
i=1,2$. Then for any functions $x\in V\mapsto
z_x=(z_{1,x},z_{2,x})$, satisfying the relation
$$z_{i,x}=\lambda \prod_{y\in S(x)}{a_{i0}+
a_{i1}z_{1,y}+a_{i2}z_{2,y}\over
a_{00}+a_{01}z_{1,y}+a_{02}z_{2,y}}, \ \ i=1,2,$$
there exists a unique $G$-HC-Gibbs measure $\mu$, and vice versa.
We consider the translation-invariant
solutions such that $z_x=z\in
R^2_+$, $x\neq x_0$.

By works \cite{RSh} and \cite{RKh1} for $k=2$ it is know next theorems.

\begin{thm}\cite{RSh} Let $k=2$ and $\lambda _{cr} =\frac{9}{4}$. Then for HC model in the case $G=\textit{hinge}$  there exists a unique translation-invariant Gibbs measure $\mu _{0}$ for $\lambda \le \lambda _{cr} $, and there exist only three translation-invariant Gibbs measures $\mu _{0} ,\, \, \mu _{1} ,\, \, \mu _{2}$ for $\lambda >\lambda _{cr}$.
\end{thm}

\begin{thm}\cite{RSh} Let $k=2$ and $\lambda _{cr}^{*} =1$. Then for HC model in the case $G=\textit{wand}$ there exists only one translation-invariant Gibbs measure $\mu _{0}^{*}$ for $\lambda \le \lambda _{cr}^{*}$ and there exist only three translation-invariant Gibbs measures $\mu _{0}^{*} ,\, \, \mu _{1}^{*} ,\, \, \mu _{2}^{*}$ for $\lambda >\lambda _{cr}^{*}$.
\end{thm}

\begin{thm} \cite{RKh1} Let $k=2$ and $\lambda _{0} \approx 7,0355$. Then for HC model in the case $G=\textit{hinge}$ the following statements hold:

1. The measure $\mu _{0}$ is non-extreme for $\lambda >\lambda _{0}$.

2. The measure $\mu _{0} $ is extreme for $0<\lambda <\lambda _{0} $ and measures $\mu _{1} ,\mu _{2} $ are extreme for $2.25<\lambda <0.5\cdot (5\sqrt{2}-1)$.
\end{thm}

\begin{thm} \cite{RKh1} Let $k=2$, $\lambda _{1} \approx 2.287572$ and $\lambda _{2} \approx 1.303094$. Then for HC model in the case $G=\textit{wand}$ the following statements hold:

1. The measure $\mu _{0}^{*} $ is non-extreme for $\lambda >\lambda _{1}$.

2. The measure $\mu _{0}^{*}$ is extreme for $0<\lambda <\lambda _{1}$ and measures $\mu _{1}^{*} ,\, \, \mu _{2}^{*}$ are extreme for $1<\lambda <\lambda _{2}$.
\end{thm}

The next theorems hold.

\begin{thm} If $k=2$ then for HC model in the case $G=\textit{hinge}$ there exist at least two extreme Gibbs measures for $0.5\cdot (5\sqrt{2} -1)<\lambda <\lambda _{0}$.
\end{thm}

\textbf{Proof.} Let $k=2$. By theorem 4 it is know that there exists a unique translation-invariant Gibbs measure $\mu _{0}$ for $0<\lambda \le \lambda _{cr}$. By theorem 6 the measure $\mu _{0}$ is extreme for $0<\lambda <\lambda _{0}$. For $\lambda >\lambda _{cr} =2,25$ we have the measure $\mu _{0}$ and at last two new measures mentioned in Theorem 4. If we assume that all the new measures are not extreme in $(0.5\cdot (5\sqrt{2} -1),\lambda _{0} )$ then it remains only one extreme measure $\mu _{0}$. But in this case the non-extreme
measures can not be decomposed only into the unique measure $\mu _{0} $. Consequently, at least
one of the new measures for $0.5\cdot (5\sqrt{2} -1)<\lambda <\lambda _{0}$ must be extreme or decomposable into other extreme measures. The theorem is proved.

\begin{thm} If $k=2$ then for HC model in the case $G=\textit{wand}$ there exist at least two extreme Gibbs measures for $\lambda _{2} <\lambda <\lambda _{1}$.
\end{thm}

\textbf{Proof.} It is proved similarly to the previous theorem.\

\textbf{Usefulness:} The main problem in the theory of Gibbs measures is the full description of the set of all limit Gibbs measures, but this problem is not solved fully. In \cite{KhR} for HC model nonuniqueness of weakly periodic Gibbs measures was proved. In this work we found new measures (see Theorem 2, part 1) which are different from the know old measures (see Theorem 1, parts 2,3). Every new measure which is we found expands the set of known Gibbs measures and each of them gives a new phase for the physical system. Moreover, definition of all extreme Gibbs measures for given Hamiltonian give the full description of all limit Gibbs measures.\

\textbf{Acknowledgments.} The authors are very grateful to Professor U. A. Rozikov for his useful discussions.

\end{document}